\documentclass[aps,prl,reprint,a4paper,preprintnumbers,superscriptaddress,longbibliography,nofootinbib,amsmath,amssymb,floatfix]{revtex4-2}
\pdfoutput=1
\usepackage{graphicx}
\usepackage{scalefnt}
\usepackage[dvipsnames]{xcolor}
\usepackage[acronym]{glossaries}
\let\oldglsentryshort\glsentryshort
\usepackage{orcidlink,hyperref}
\hypersetup{colorlinks=true,linkcolor=blue,citecolor=blue,urlcolor=blue}
\graphicspath{{../}{./}}
\usepackage[capitalize]{cleveref}

\newcounter{notecount}

\newcommand{\abbrev}[1]{{\scalefont{.9}#1}}
\renewcommand{\glsentryshort}[1]{\abbrev{\oldglsentryshort{#1}}}
\newcommand{\myacrodef}[3]{\newacronym{#2}{#2}{#3}\newcommand{#1}{\gls{#2}}}
\myacrodef{\dwf}{DWF}{domain-wall fermion}
\myacrodef{\uv}{UV}{ultraviolet}
\myacrodef{\lft}{LFT}{lattice field theory}
\myacrodef{\via}{VIA}{vacuum-insertion approximation}
\myacrodef{\ckm}{CKM}{Cabbibo-Kobayashi-Maskawa}
\myacrodef{\rg}{RG}{renormalization group}
\myacrodef{\sftx}{SFTX}{short flow-time expansion}
\newacronym[plural=effective field theories,shortplural=EFTs]{EFT}%
{EFT}{effective field theory}

\myacrodef{\gf}{GF}{gradient flow}
\myacrodef{\qft}{QFT}{quantum field theory}
\newcommand{\qcd}{\abbrev{QCD}}
\myacrodef{\lhc}{LHC}{Large Hadron Collider}
\myacrodef{\lo}{LO}{leading order}
\myacrodef{\nlo}{NLO}{next-to-leading order}
\myacrodef{\nnlo}{NNLO}{next-to-next-to-leading order}
\myacrodef{\pdf}{PDF}{parton density function}
\myacrodef{\sm}{SM}{standard model}
\myacrodef{\bsm}{BSM}{beyond-the-\gls{SM}}
\myacrodef{\hqe}{HQE}{heavy-quark expansion}
\myacrodef{\hqet}{HQET}{heavy-quark effective theory}
\newcommand{\alphas}{\alpha_\text{s}}

\newcommand{\two}{two}
\newcommand{\three}{three}
\newcommand{\four}{four}
\newcommand{\gev}{\,\text{Ge\hspace{-0.1em}V}}
\newcommand{\mev}{\,\text{Me\hspace{-0.1em}V}}

\newcommand{\MSbar}{\ensuremath{\overline{\abbrev{\text{MS}}}}}

\begin{document}

\preprint{SI-HEP-2026-07,~~P3H-26-023,~~ZU-TH 13/26,~~TTK-26-06}

\title{Bag Parameters for Heavy Meson Lifetimes from a Lattice QCD Calculation at the SU(3)$_{\rm F}$-symmetric Point}

\author{Matthew Black\orcidlink{0000-0002-8952-1755}}
\email{matthew.black@ed.ac.uk}
\affiliation{Higgs Centre for Theoretical Physics, School of Physics and Astronomy, University of Edinburgh, Edinburgh EH9 3JZ, United Kingdom}
\affiliation{Theoretische Physik 1, Center for Particle Physics Siegen, Naturwissenschaftlich-Technische Fakult\"at, Universit\"at Siegen, 57068 Siegen, Germany}

\author{Robert V.~Harlander\orcidlink{0000-0002-8715-2458}}
\affiliation{Institute for Theoretical Particle Physics and Cosmology, RWTH Aachen University, 52056 Aachen, Germany}

\author{Jonas T. Kohnen\orcidlink{0009-0001-2904-5945}}
\affiliation{Institute for Theoretical Particle Physics and Cosmology, RWTH Aachen University, 52056 Aachen, Germany}

\author{\\Fabian Lange\orcidlink{0000-0001-8531-5148}}
\affiliation{Physik-Institut, Universität Zürich, Winterthurerstrasse 190, 8057 Zürich, Switzerland}
\affiliation{PSI Center for Neutron and Muon Sciences, 5232 Villigen PSI, Switzerland}

\author{Antonio Rago\orcidlink{0000-0003-1192-5538}}
\affiliation{IMADA and Quantum Theory Center, University of Southern Denmark, Odense, Denmark}

\author{Andrea Shindler\orcidlink{0000-0003-3693-8300}}
\affiliation{Institute for Theoretical Particle Physics and Cosmology, RWTH Aachen University, 52056 Aachen, Germany}
\affiliation{Nuclear Science Division, Lawrence Berkeley National Laboratory, Berkeley, California 94720, USA}
\affiliation{Department of Physics, University of California, Berkeley, CA 94720, USA}

\author{Oliver Witzel\orcidlink{0000-0003-2627-3763}}
\affiliation{Theoretische Physik 1, Center for Particle Physics Siegen, Naturwissenschaftlich-Technische Fakult\"at, and Cluster of Excellence ``Color meets Flavor'', Universit\"at Siegen, 57068 Siegen, Germany}

\date{\today}

\begin{abstract}
We calculate the dimension-six $\Delta Q=0$ four-quark matrix elements describing heavy-meson lifetime ratios using the gradient flow with its \sftx{} as a renormalization procedure.
On six \abbrev{RBC/UKQCD} 2+1-flavor domain-wall fermion ensembles, 
we determine flowed bag parameters for physical charm and strange quarks and match to the \MSbar\ scheme with perturbative \sftx{} coefficients through \nnlo. 
A multiscale matching procedure using renormalization-group running improves the extrapolation to zero flow time. 
For the operators relevant to $\tau(D_s)/\tau(D^0)$ at the SU(3)$_{\rm F}$ symmetric
point, we obtain $B_1^{\MSbar}(3\gev)=1.0524(97)$,
$B_2^{\MSbar}(3\gev)=0.9621(70)$, $\epsilon_1^{\MSbar}(3\gev)=-0.2275(76)$, and
$\epsilon_2^{\MSbar}(3\gev)=-0.0005(8)$ 
using a specific choice of evanescent operators.
This is the first lattice-\qcd{} determination of $\Delta Q=0$ four-quark operators
with a full error budget. 
It opens the path toward higher-precision
predictions of heavy-meson lifetimes and similar quantities exhibiting operator
mixing under renormalization. 
\end{abstract}
\maketitle

\paragraph{Introduction ---}

The physics of heavy hadrons provides insight into several crucial aspects of
the \sm. While the binding of the quarks is governed by strong interactions,
mixing and decay of the hadrons is induced by weak interactions, with strong
sensitivity to the \ckm\ matrix elements.  The comparison of precise
measurements and theoretical predictions of the heavy-hadron properties thus
provides crucial input for the indirect search of physics beyond the \sm.  A
summary of the theoretical and experimental status can be found in
Refs.~\cite{HFLAV:2024ctg,PDG:2024cfk,Albrecht:2024oyn}.

Theoretical predictions of heavy-hadron properties need to account for both
perturbative (short-distance) and nonperturbative (long-distance) effects.
A suitable framework for this is provided by effective field theories, where the
perturbative effects are captured by the Wilson coefficients, while the
determination of the associated matrix elements requires nonperturbative
methods.

To date, sum-rule techniques have been used to estimate these nonperturbative
effects for lifetimes and lifetime ratios in the framework of
\hqet~\cite{Kirk:2017juj,Kirk:2018kib,King:2021xqp,
King:2021jsq,King:Thesis22,Black:2024bus}.  \Gls{LFT}\ calculations have so far
delivered phenomenologically relevant results for short-distance contributions
to heavy-meson
mixing~\cite{Carrasco:2014uya,Carrasco:2015pra,Bazavov:2017weg,Carrasco:2013zta,Aoki:2014nga,Gamiz:2009ku,Bazavov:2016nty,Dowdall:2019bea,Boyle:2021kqn},
but, despite similarities in the calculation and early work \cite{DiPierro:1998ty,DiPierro:1999tb,Becirevic:2001fy}, \lft\ is yet to
provide the nonperturbative input for heavy-meson lifetimes with full error
budgets.
An alternative approach using HQET has been proposed in Ref.~\cite{Lin:2022fun}.

In this Letter, we consider $D_s$ mesons, which constitute an ideal laboratory
for establishing and validating our method to nonperturbatively renormalize the
matrix elements on the lattice.  The gradient flow
is used for the operator renormalization and enables us to take the
continuum limit. Subsequently, we convert the results to the \MSbar\ scheme using
perturbatively-evaluated matching coefficients. 
In this proof-of-principle calculation, we focus on lifetime ratios in the limit of SU(3)-flavor symmetry between $u$, $d$, and $s$ quarks, which provide clean probes of the leading spectator effects in the \hqe.
In these ratios, contributions due to so-called ``eye diagrams'' and mixing with lower-dimensional operators cancel up to SU(3)$_\mathrm{F}$-breaking effects.
The effect of eye diagrams was found to be $O(0.2\%)$ in \hqet\ using the sum-rules approach, and the strange quark-mass corrections were found to be $O(0.1\%)$~\cite{King:2021jsq,King:Thesis22}.
The error budget quoted in this work applies to the bag parameters computed in this SU(3)$_\mathrm{F}$-symmetric setup, while a direct lattice calculation of the corresponding SU(3)$_\mathrm{F}$-breaking corrections is left to future work.

Nevertheless, our approach is applicable also to the calculation of absolute
lifetimes in principle, even if some of the steps may require additional efforts
in order to reach similar precision. The analogous calculation for the simpler
case of heavy-meson mixing and further details on the lifetime calculation are
presented in the companion paper~\cite{Black:2026dzp}, while preliminary results have been presented in Refs.~\cite{Black:2023vju,Black:Thesis24,Black:2024iwb}.

\begin{figure}[tb]
  \begin{center}
    \begin{tabular}{cc}
      \includegraphics[width=.5\columnwidth]{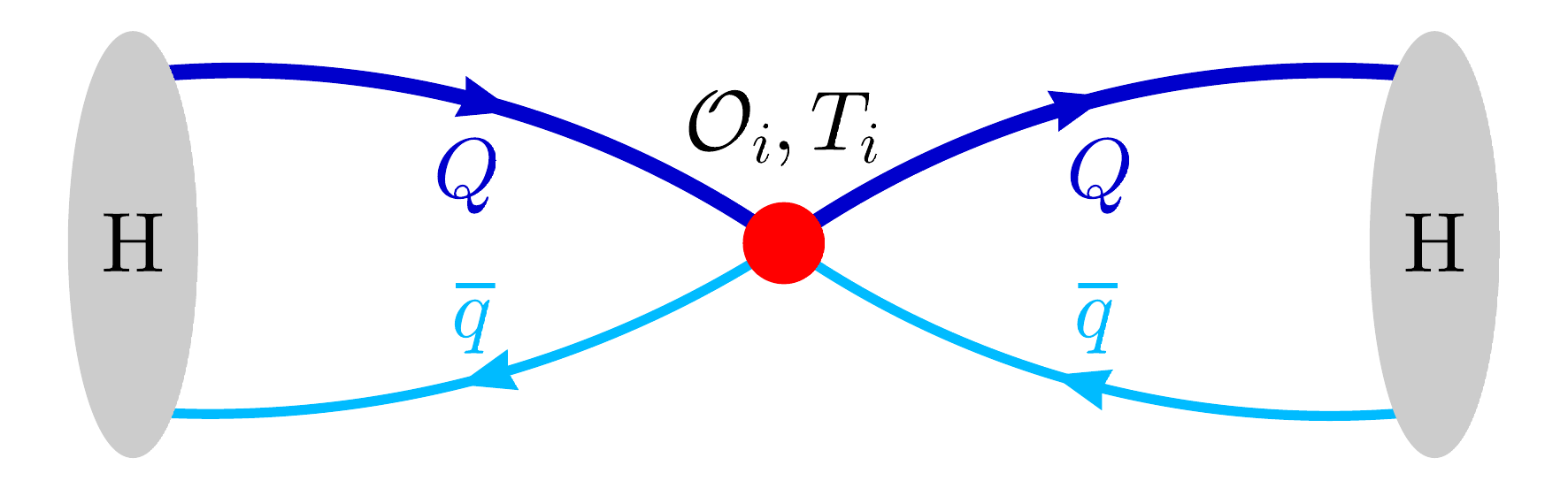} &
      \includegraphics[width=.5\columnwidth]{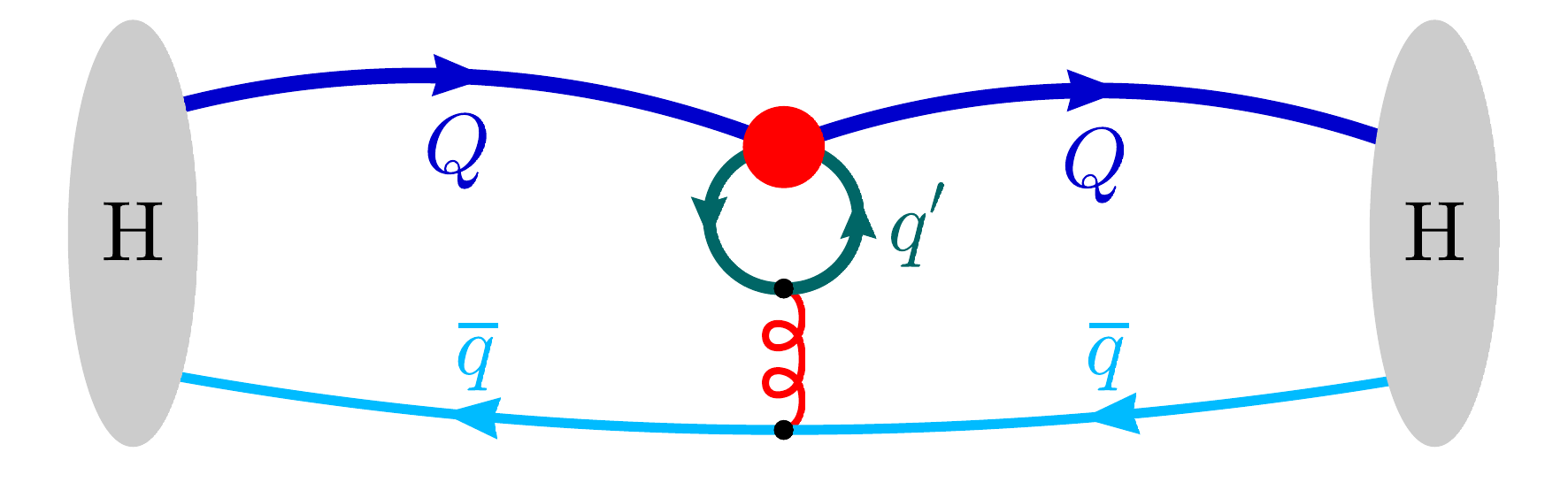}\\
      (a) & (b) 
    \end{tabular}
    \caption{\label{fig:lifetimes} (a) \lo\ diagram for the four-quark operator
    contribution to heavy-meson decay in \gls{HQE}; (b)~eye diagram (not
    considered in this Letter). The diagrams were produced with the help of \textsc{FeynGame}~\cite{Harlander:2020cyh,Bundgen:2025utt}.}
  \end{center}
\end{figure}

\paragraph{Operators and bag parameters ---}

\glsreset{HQE} Within the \hqe~\cite{Nikolaev:1973uu,Neubert:1993mb}, the lifetime of a heavy hadron $H$ is
governed by the imaginary part of the forward $H\to H$ matrix elements of local $\Delta Q=0$ operators. 
The leading spectator effects incorporating the meson's light valence quark arise at order $1/m_Q^3$ 
and are given by the following dimension-six four-quark operators (see e.g.~Refs.~\cite{Kirk:2017juj,Piscopo:2021ogu}):
\begin{equation}\label{eq:4quark}
  \begin{aligned}
    {\cal O}_1 &= (\bar{Q}\gamma_\mu (1-\gamma_5) q)
    (\bar{q}\gamma^\mu (1-\gamma_5) Q)\,, \\
  {\cal O}_2 &= (\bar{Q} (1-\gamma_5) q)(\bar{q} (1+\gamma_5) Q)\,, \\
  T_1 &= (\bar{Q}\gamma_\mu (1-\gamma_5) T^Aq)(\bar{q}\gamma^\mu (1-\gamma_5)
  T^AQ)\,, \\
  T_2 &= (\bar{Q} (1-\gamma_5) T^Aq)(\bar{q} (1+\gamma_5) T^AQ)\,,
  \end{aligned}
\end{equation}
where the $T^A$ are the generators of SU(3) color. A \lo\ Feynman diagram
contributing to the forward matrix elements of these operators is shown in
\cref{fig:lifetimes}\,(a). These matrix elements are parametrized as
\begin{equation}\label{eq:DB0_ME}
  \begin{aligned}
  \langle H|{\cal O}_1|H\rangle &=
  f_{H}^2\,M_{H}^2\,{B}^{H}_1\,, \\
  \langle H|{\cal
    O}_2|H\rangle &= \frac{M_{H}^2}{(m_Q+m_q)^2}f_{H}^2\,M_{H}^2\,{B}_2^{H}\,,
   \\ \langle H|T_1|H\rangle &= f_{H}^2\,M_{H}^2\,\epsilon_1^{H}\,,
  \\ \langle H|T_2|H\rangle &=
  \frac{M_{H}^2}{(m_Q+m_q)^2}f_{H}^2\,M_{H}^2\,\epsilon_2^{H}\,,
  \end{aligned}
\end{equation}
with $M_{H}$ the mass of the heavy meson $H$.  The bag parameters $B_1^H$,
$B_2^H$, $\epsilon_1^H$, and $\epsilon_2^H$ as well as the decay constant $f_H$
encode contributions due to strong interactions whose determinations require
nonperturbative methods.  In this Letter, we focus on a bound charm-strange state, i.e.\
$H=D_s$. We expect our method to apply also to other $D$ or $B$ mesons;
we leave this for future work.

Aside from diagrams like \cref{fig:lifetimes}\,(a), the operators in
\cref{eq:4quark} also lead to diagrams where the light valence quark of the meson
is not directly connected to the composite operator, sometimes called
``eye diagrams''; see \cref{fig:lifetimes}\,(b). 
Their consistent treatment requires an extension of the operator basis in \cref{eq:4quark}. 
In a first step toward the full calculation of heavy-meson lifetimes, we determine the matrix elements relevant to \textit{SU(3)$_\mathrm{F}$-symmetric ratios} of lifetimes like $\tau(D_s)/\tau(D^0)$, where these eye diagrams cancel. 

\begin{figure}[t]
    \includegraphics[width=0.7\columnwidth]{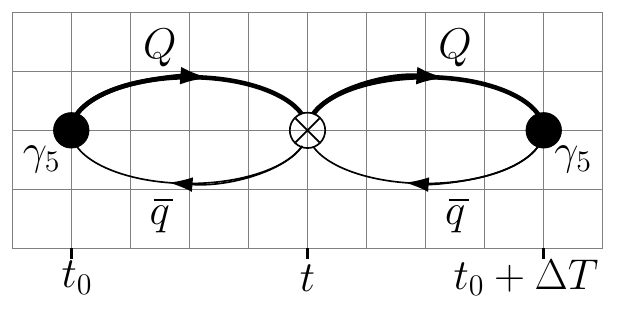}
    \caption{Quark-line diagram for the lattice calculation of $\Delta Q=0$ operators.}
    \label{fig:QuarkLineDiagram}
\end{figure}

\paragraph{Lattice calculation ---}
\Cref{fig:QuarkLineDiagram} illustrates the calculation of a \three-point
function on the lattice. Mesons are created at Euclidean times $t_0$ and
$t_0+\Delta T$.  The four-quark interactions described by the operators defined
in \cref{eq:4quark} are contracted at $t\in[t_0,t_0+\Delta T]$.  By placing the
sources at equal time separations, propagators can be reused to calculate
\three-point functions with different separation $\Delta T$ concurrently
(cf.~Ref.~\cite{Boyle:2018knm}).

Adjusting the spin-color algebra factors in the ratios seen in Ref.~\cite{Boyle:2018knm} to the $\Delta Q=0$ case, the bag parameters can be obtained from the following ratio
\begin{align}
    R_X(t_0,t,\Delta T) &= \frac{C^{\rm 3pt}_{X}(t_0,t,t_0+\Delta
      T)}{C_X(t_0,t)\,C_X(t_0+\Delta T,t)}
    \to B_X\,,
    \label{eq.ratio}
\end{align}
where the arrow indicates the limits $t_0\ll t\ll t_0+\Delta T$ and $\Delta T\to\infty$.
The \two- and \three-point functions are given by
\begin{equation}\label{eq::acis}
  \begin{aligned}
    C^\text{3pt}_{X}(t_1,t_2,t_3) &=
    \langle0| j_P(t_1)X(t_2)j_P^\dagger(t_3)|0\rangle\,,\\
    C_X(t_1,t_2) &=
    \langle0| j_X(t_2)j_P^\dagger(t_1)|0\rangle\,,
  \end{aligned}
\end{equation}
where $t_1<t_2<t_3$ are generic positions, and $j_P=\bar{q}\gamma_5 Q$ and $j_X=\bar{q}\Gamma_X Q$
are zero-momentum currents. Note that we are using $X$ both as an operator and as an index in \cref{eq::acis}.
In order to have appropriate overlap with the bag parameters, 
$\Gamma_{\mathcal{O}_1}=\Gamma_{T_1}=\gamma_0\gamma_5$, and
$\Gamma_{\mathcal{O}_2}=\Gamma_{T_2}=\gamma_5$.

\paragraph{Gradient flow ---}
Bare correlators as well as the ratio of \cref{eq.ratio} are
\uv\ divergent. We obtain the \MSbar-renormalized hadronic matrix elements by
closely following the method of
Refs.\,\cite{Suzuki:2013gza,Makino:2014taa,Taniguchi:2020mgg,Suzuki:2021tlr,Black:2025gft,Francis:2025pgf,Francis:2025rya}
which is based on the
\gf~\cite{Narayanan:2006rf,Luscher:2009eq,Luscher:2010iy,Luscher:2013cpa} and
the \sftx~\cite{Luscher:2011bx,Suzuki:2013gza,Luscher:2013vga}.  To be
specific, we replace the operators $X$ and the currents $j_X$ in~\cref{eq::acis} by their flowed versions $\widetilde{X}(\tau)$ and $\widetilde{j}_X(\tau)$, i.e. the fields in these quantities are evolved according to
\begin{equation}\label{eq::farc}
  \begin{aligned}
    \partial_\tau \mathcal{A}_\mu(\tau) &= D_\nu(\tau)G_{\nu\mu}(\tau)\,,&&&
    \mathcal{A}_\mu(\tau=0)&=A_\mu\,,\\ \partial_\tau \chi(\tau) &= D^2(\tau)
    \chi(\tau)\,,&&& \chi(\tau=0)&=\psi\,,
  \end{aligned}
\end{equation}
with $\tau$ indicating the flow time.
$\chi(\tau)$ represents the quark fields, $G_{\mu\nu}$ is the field-strength tensor, and $D_\mu$ the covariant derivative, each with the regular
gluon field $A_\mu$ replaced by its flowed version $\mathcal{A}_\mu(\tau)$.
On the lattice, we use the Wilson kernel to evolve the gauge fields, and apply the
lattice Laplace operator to the propagators~\cite{Luscher:2013vga}.  
For finite $\tau >0$, the \gf\ regularizes
the \uv\ divergences, and the continuum limit
combining results from the
different ensembles can be taken. Power-divergent operator mixing on the lattice is absent in this approach and shifted to the \sftx. Note that any wave function
renormalizations for the flowed and unflowed quark fields cancel in the ratio of
\cref{eq.ratio}.

At this point, the continuum bag parameters depend on the flow
time $\tau$; we may consider them as renormalized in the \gf\ scheme 
denoting them by $\widetilde{B}_X(\tau)$. They can be converted to the \MSbar\ scheme
at a scale $\mu$ as follows. Using the
\sftx~\cite{Luscher:2011bx,Suzuki:2013gza,Luscher:2013vga}, the \gf-renormalized
bag parameters are multiplied by the matching
matrices,
\begin{equation}\label{eq::kava}
  \begin{aligned}
    \begin{pmatrix}
      B_i(\tau,\mu)\\
      \epsilon_i(\tau,\mu)
    \end{pmatrix}
    =
    \zeta^{-1}_{B,i}(\tau,\mu)
    \begin{pmatrix}
      \widetilde{B}_i(\tau)\\
      \widetilde{\epsilon}_i(\tau)
    \end{pmatrix}\,,
  \end{aligned}
\end{equation}
for $i\in\{1,2\}$. The matching matrices are given by
\begin{equation}\label{eq::anga}
  \begin{aligned}
    \zeta_{B,1} = \zeta_1\zeta_A^{-2}\,,\qquad
    \zeta_{B,2} = \zeta_2\zeta_P^{-2}\,,
  \end{aligned}
\end{equation} 
where $\zeta_P$ and $\zeta_A$ are the \sftx\ matching coefficients for the
pseudoscalar and the axial-vector current, and the $\zeta_{i}$ are the matching
matrices of the \four-quark operators $\{\mathcal{O}_i,{T}_i\}$ in
\cref{eq:4quark}.  They can be calculated perturbatively and are known through
\nnlo\ in \qcd~\cite{Harlander:2022tgk,Borgulat:2023xml,Black:2026dzp}.
The quantities on the l.h.s.\ of \cref{eq::kava} are related to the \MSbar\ bag parameters by taking the
$\tau\to0$ extrapolation, i.e.
\begin{equation}\label{eq:limtau0}
    B_X^{\MSbar}(\mu) = \lim_{\tau\to0} B_X(\tau,\mu).
\end{equation}
This limit is non-trivial and involves identifying a suitable
region from which such an extrapolation can be performed as we discuss later.

\paragraph{Numerical analysis ---}
Using the six \abbrev{RBC/UKQCD} $2+1$-flavor Shamir \dwf\ and Iwasaki gauge field
ensembles \cite{Allton:2008pn,Aoki:2010dy,Blum:2014tka,Boyle:2018knm} listed
in \cref{tab:confs}, we create multiple $\mathbb{Z}_2$ wall sources
\cite{McNeile:2006bz,Boyle:2008rh} per gauge-field configuration.  For strange
quarks, we apply Gaussian smearing~\cite{Gusken:1989ad} to the source before
inverting a Shamir \dwf~\cite{Kaplan:1992bt,Shamir:1993zy,Furman:1994ky},
whereas for charm quarks we directly invert a heavy, stout-smeared M\"obius
\dwf~\cite{Morningstar:2003gk,Brower:2012vk,Cho:2015ffa}.  In both cases, the
mass is tuned close to its physical value.

Evolving the gauge field and quark propagators along the \gf\ requires solving
the ordinary differential equations of \cref{eq::farc}. We implement them
using a third-order Runge-Kutta algorithm in the \texttt{Grid}~\cite{Grid16}
and \texttt{Hadrons}~\cite{HADRONS,Hadrons22} lattice-\qcd\ software
libraries.
The \gf\ evolution runs with step size $\delta\tau/a^2 = 0.01$, with
measurements of the matrix elements carried out every 10~steps
(increasing to 40~steps toward larger $\tau$).

For each value of the flow time, we perform combined correlated fits to~\cref{eq.ratio} for a range of values of $\Delta T$ where excited states are not statistically resolved and consistent plateaus can be identified; the minimum $\Delta T$ considered on each ensemble is listed in~\cref{tab:confs}. Details of the numerical analysis are discussed in Ref.~\cite{Black:2026dzp}.

\begin{figure*}[tb]
    \includegraphics[width=0.47\textwidth]{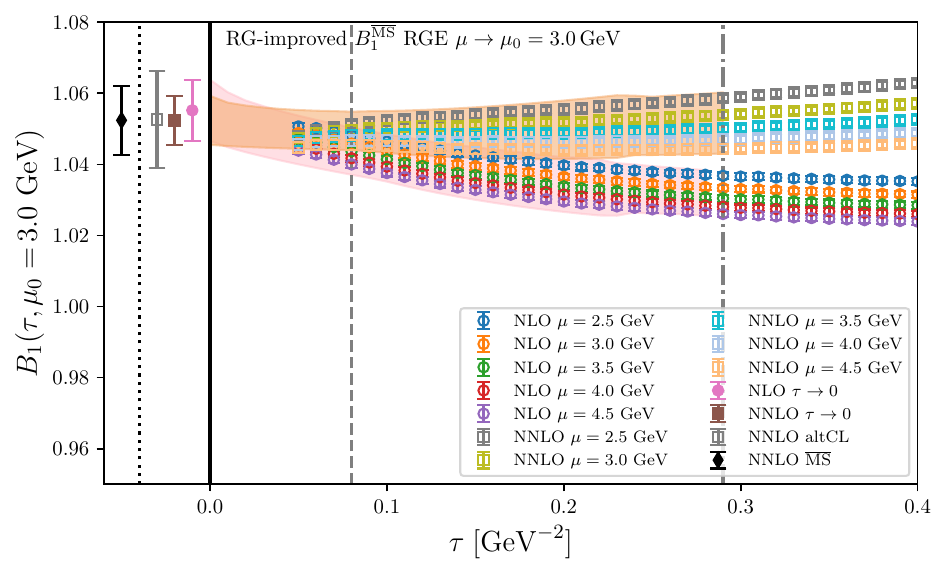}\hfill
    \includegraphics[width=0.47\textwidth]{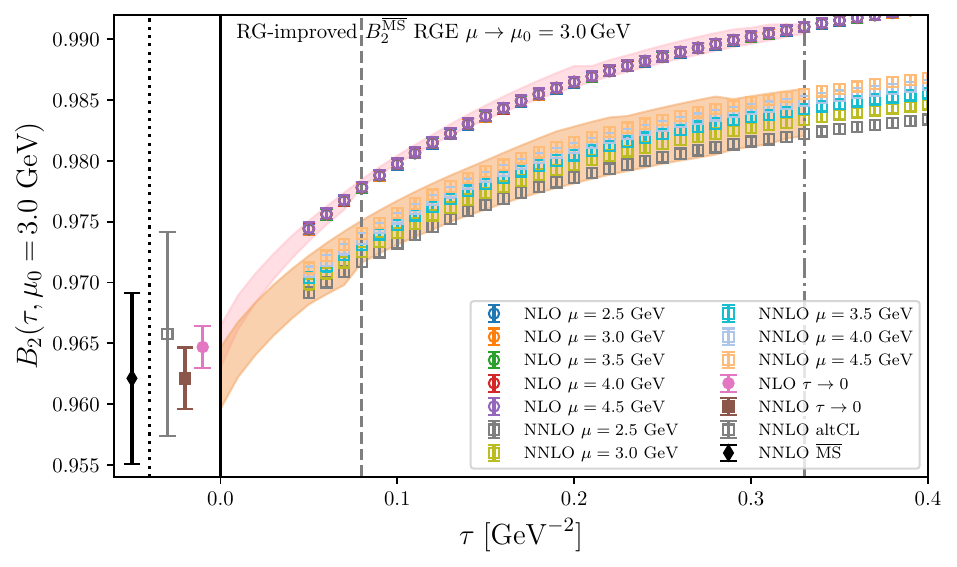}\\
    \includegraphics[width=0.47\textwidth]{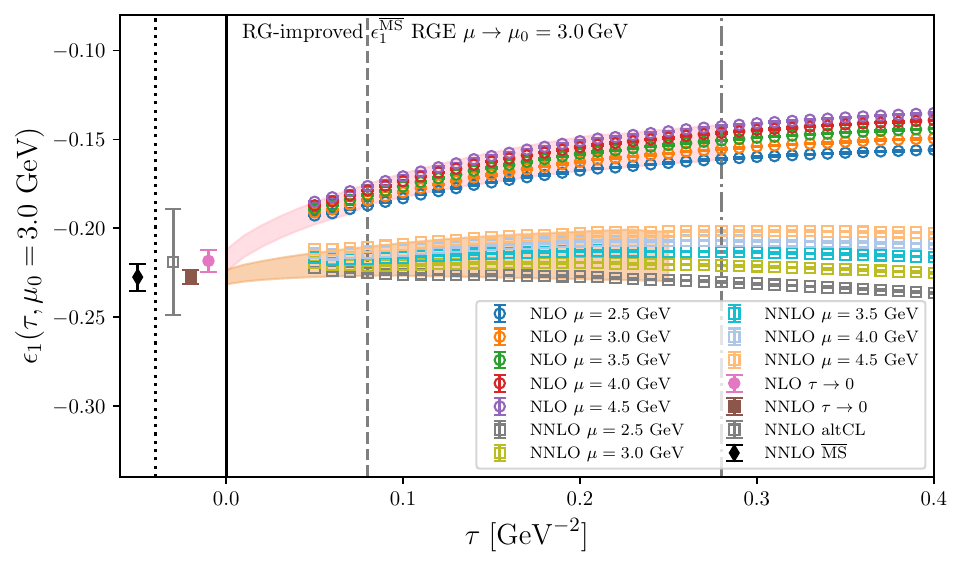}\hfill
    \includegraphics[width=0.47\textwidth]{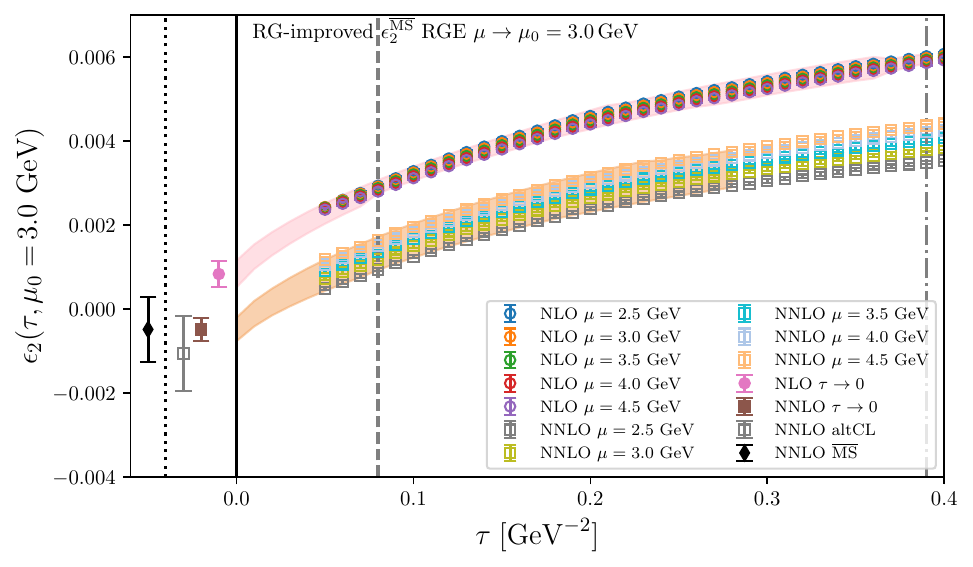}
    \caption{Zero-flow-time extrapolation of the $\Delta Q=0$ bag parameters -- $B_1$ (top left), $B_2$ (top right), $\epsilon_1$ (bottom left), $\epsilon_2$ (bottom right). The matching to \MSbar\ is performed at both NLO and NNLO using $\mu\in\{2.5,3.0,3.5,4.0,4.5\}\gev$ before running back to $\mu_0=3\gev$ and uses the flow-time \rg\ equation to set flow-time logarithms to zero and improve the extrapolations. The pink and orange shaded areas indicate the spread of all 
    fits with $p$-value $> 0.05$ at NLO and NNLO, resulting in the pink and brown data points in the left panel respectively. 
    The gray data points estimate systematic effects due to discarding our coarsest ensembles; the black data points show our final result with all errors added in quadrature.
    The dashed lines indicate
    $\tau_{\rm min}=0.08\gev^{-2}$, whereas 
    dash-dotted lines 
    mark the maximum flow time included in our fit results.
    }
    \label{fig:lattice+sftx}
    \vspace{-3pt}
\end{figure*}

Since the gradient flow is only effective
if the ``smearing radius'' $\sqrt{8\tau}$ is sufficiently large in relation to
the lattice spacing, we only include data in our analysis with
$\tau>\tau_\text{min}$, where $8\tau_\text{min}=2a^2_{\rm max}$.
Using the coarsest lattice spacing of our dataset, this translates to $\tau_\text{min} = 0.08\gev^{-2}$.
Using ${\cal O}(a)$-improved actions, we take the continuum limit using a linear ansatz in $a^2$ and also parametrize sea-pion mass effects using
\begin{equation}
    \begin{aligned}
        \widetilde{B}_i^{\rm latt}(a^2,M_\pi^{\rm latt};\tau) = &\widetilde{B}_i(\tau) + C\,a^2 \\
        +& D\,a^2\,\Big[(M_\pi^{\rm latt})^2 - (M_\pi^{\rm phys})^2\Big],
    \end{aligned}
\end{equation}
where $\widetilde{B}_i(\tau)$ denotes the continuum-extrapolated value at physical pion mass~\cite{PDG:2024cfk}, while $C$ and $D$ are fit parameters describing the leading discretization and sea-pion effects. 
Estimates of systematic uncertainties are discussed in Ref.~\cite{Black:2026dzp}.

\paragraph{Zero-flow-time limit ---}
In the final step of our procedure, the flowed bag parameters are multiplied by
the inverse matching matrices defined in \cref{eq::kava,eq::anga}, and the
result is extrapolated to $\tau=0$, resulting in the bag parameters in the
\MSbar\ scheme.  

To derive the bag parameters in the \MSbar\ scheme at a
certain value of $\mu=\mu_0$, we can either insert the (inverse)
matching matrices at this value, $\zeta^{-1}(\tau,\mu_0)$, into \cref{eq::kava},
or evaluate them at a different scale $\mu$, and subsequently evolve them
to $\mu_0$ according to regular \rg\ evolution~\cite{Harlander:2020duo},
\begin{equation}\label{eq::blow}
  \begin{aligned}
    \mu^2\frac{\rm d}{{\rm d}\mu^2}\zeta(\tau,\mu) = \zeta(\tau,\mu)\gamma(\alphas(\mu))\,,
  \end{aligned}
\end{equation}
where $\gamma$ is the regular (unflowed) anomalous dimension of
$R_X$~\cite{Buras:1989xd,Ciuchini:1997bw,Buras:2000if,Aebischer:2025hsx}.
We denote the resulting matching matrix as $\zeta(\tau,\mu\to\mu_0)$. 
Different initial values of $\mu$ typically change the $\tau$ dependence of
$\zeta^{-1}(\tau,\mu\to\mu_0)\widetilde{B}(\tau)$, and therefore lead to
small variations in the
$\tau\to0$ limit, which we will use to estimate part of the systematic
uncertainty. 
The value of $\alphas(\mu)$ is obtained from our reference
value $\alphas(3\,\gev)=0.2535$ by \rg\ evolution to the corresponding
perturbative order with the help of \textsc{RunDec}\,\cite{Chetyrkin:2000yt,Herren:2017osy}.

Furthermore, we can improve the perturbative convergence of our matching procedure by resumming the logarithmic $\tau$ dependence in the matching coefficients via the flow-time \rg\ equation~\cite{Harlander:2020duo}
\begin{equation}\label{eq:flowRGE}
    \tau\frac{{\rm d}}{{\rm d}\tau}\tilde{\mathcal{O}}(\tau) = \tilde{\gamma}\tilde{\mathcal{O}}(\tau)\,,\quad\text{with}\quad\tilde{\gamma}=\left(\tau\frac{{\rm d}}{{\rm d}\tau}\zeta\right)\zeta^{-1}\,,
\end{equation}
where the flowed anomalous dimensions $\tilde{\gamma}$ are provided in Ref.~\cite{Black:2026dzp}.
This allows us to run our data for the bag parameters between different flow
times; in particular, we can evolve all our data from $\tau$ to a specific flow
time $\tau_0$ where higher-order corrections to the matching matrix $\zeta$ are
particularly small.
Specifically, we choose
$\tau_0={\rm e}^{-\gamma_\text{E}}/(2\mu_0^2)$~\cite{Harlander:2018zpi}, where
$\gamma_\text{E}=0.5772\ldots$ is the Euler-Mascheroni constant.

Our final results for $B^{\MSbar} _i(\mu_0)$ and $\epsilon_i^{\MSbar}(\mu_0)$
are obtained by performing a single global fit to each
$B(\mu\to\mu_0,\tau\to\tau_0)$ for $\mu\in\{2.5,3.0,3.5,4.0,4.5\}\gev$ and $\mu_0=3\gev$, with the
functional form
\begin{equation}\label{eq::june}
  \begin{aligned}
    f(\tau,\mu) = c_0 +
    \tau \left(c_1^{(\mu)} + c_2^{(\mu)}\ln (\mu^2\tau)\right)\,,
  \end{aligned}
\end{equation}
where $c_0$ is a common parameter to all fit functions, and $c_{1,2}^{(\mu)}$
are independent for each $\mu$ used.  By constraining $c_0$ to be global to all
fit functions, we ensure a solution for which the perturbative \rg\ evolution is
consistent across the chosen range of $\mu$.
We evaluate fits within all possible intervals from $\tau_\text{min}$
up to at most $\tau_\text{max}=0.4\,\gev^{-2}$,  
and identify the spread of all fit results for $c_0$ with $p$-value $>0.05$ as
the symmetrized uncertainty interval for the bag parameters at scale $\mu_0$.
This interval accounts for both statistical and systematic uncertainties of
the $\tau\to0$ extrapolation.

\paragraph{Results ---}

\Cref{fig:lattice+sftx} shows the \sftx\ and $\tau\to0$ limits for the bag
parameters of the $\Delta Q=0$ operators for the lifetime ratio
$\tau(D_s)/\tau(D^0)$.  The colored bands show the spread of all fit
variations considered. The results at $\tau=0$ at \nlo\ and \nnlo\ are shown
in the left panel as pink and brown data points, respectively.  
The gray data points are based on an alternative analysis to estimate systematic effects by discarding our coarsest ensembles (C1, C2) in the continuum extrapolation. This removes 1/3 of our total dataset and an entire lattice spacing, leading to larger uncertainties since the continuum extrapolations are less constrained.
Finally, the black data points show our final results
incorporating in addition systematic uncertainties due to truncating the
perturbative series as well as estimates for discretization and quark mass
tuning uncertainties.  We find
\begin{align}
\begin{aligned}
    B_1^{\MSbar}(3\gev) &= \phantom{-}1.0524(97), \\
    B_2^{\MSbar}(3\gev) &= \phantom{-}0.9621(70),  \\
    \epsilon_1^{\MSbar}(3\gev) &= -0.2275(76),  \\
    \epsilon_2^{\MSbar}(3\gev) &= -0.0005(8). \label{eq:results}
\end{aligned}
\end{align}
for the bag parameters of lifetime operators with a
full error budget which is provided in~\cref{tab:errorbudget} and described in further detail in Ref.~\cite{Black:2026dzp}. While no recent lattice calculation exists for these quantities, they were computed
using \hqet\ sum rules with light spectator quarks in Ref.~\cite{Black:2024bus}
and then matched to \qcd\ for both $B$ and $D$ mesons. 
Accounting for different schemes of evanescent operators used (for details, see Ref.~\cite{Black:2026dzp}), we find agreement
with Ref.~\cite{Black:2024bus} 
for $B_2^{\MSbar}$ and $\epsilon_2^{\MSbar}$ within $1.5\sigma$, 
however our results
for $B_1^{\MSbar}$ and $\epsilon_1^{\MSbar}$ differ significantly. 
An improved \hqet-\qcd\ matching in the sum-rules approach that includes subleading $O(1/m_Q)$ corrections may resolve this discrepancy.
A confirmation of the increase of $\epsilon_1^{\MSbar}$
by approximately a factor of two with respect to Ref.~\cite{Black:2024bus} would be of particular phenomenological interest due
to the relatively large Wilson coefficient associated with this parameter.

The $\Delta Q=0$ operators also contribute to inclusive semileptonic $B$-meson
decays~\cite{Fael:2025xmi} and are leading uncertainties at high $q^2$.
High-precision determinations will therefore have implications to the broader
community.

\paragraph{Conclusion ---}
We have presented  the first lattice-\qcd{} determination of the SU(3)$_\mathrm{F}$-symmetric dimension-six $\Delta Q=0$
four-quark bag parameters that govern the leading spectator effects
for the lifetime ratios of heavy mesons with full error budget away from the static
limit. 
Ratios of heavy-meson lifetimes play an important role in validating the \hqe{} but also enter many phenomenological determinations such as decay-rate ratios of $b\to s\ell\ell$ currents.
A direct treatment of eye contractions, which contribute to absolute lifetimes and SU(3)$_\mathrm{F}$-breaking lifetime ratios, is the natural next step of the program.

Our result is based on \gf\ combined with the \sftx{} to obtain renormalized bag parameters in the \MSbar{} scheme.
\uv{} divergences are regularized at positive flow time such that the continuum limit can be taken without encountering power-divergent mixing, and results can be
matched to the \MSbar\ scheme using perturbatively-computed \sftx\ coefficients
(here through \nnlo). A robust $\tau\to0$ extrapolation --
restricted to a window where both discretization effects and perturbative
instabilities are under control -- completes the connection to standard
continuum phenomenology.
The determination of lifetime-ratio operators for a charm-strange system demonstrates that the \gf+\sftx\ is a viable renormalization procedure for lattice calculations, which can be further applied to complex renormalization structures including power-divergent operator mixing.
Further possible applications may include other four-quark operators exhibiting operator mixing under renormalization such as, for example, nucleon electric dipole moments.

\paragraph{Acknowledgments ---}
We thank the \abbrev{RBC/UKQCD} Collaboration for generating and making their
gauge field ensembles publicly available. Special thanks are given to Felix
Erben, Ryan Hill, and J.~Tobias Tsang for assistance in setting up the
simulation code, and we thank Anna Hasenfratz and Martin Lang for constructive
discussions.
Fruitful input and encouragement from the members of the \abbrev{CRC~TRR}~257 is
highly appreciated. Special thanks in this respect go to Alexander Lenz, Uli
Nierste, and Thomas Mannel. We are also grateful to the authors of
Ref.~\cite{Moretti:2026waq}, especially Francesco Moretti, for comparing the anomalous dimensions of the general basis of evanescent operators. 
Similarly, we thank the authors of Ref.~\cite{Aebischer:2025hsx}, especially Pol Morell, for help comparing to our anomalous dimensions. 

These computations used resources provided by the \abbrev{OMNI} cluster at the University of Siegen, the \abbrev{HAWK} cluster at the High-Performance Computing Center Stuttgart, the DeiC Large Memory HPC Type3/Hippo System managed by the eScience Center at the University of Southern Denmark, and LUMI-G at the CSC data center Finland (DeiC National HPC g.a.~DEIC-SDU-L5-13 and DEIC-SDU-N5-2024053).

This research was supported by Deutsche Forschungsgemeinschaft 
(\abbrev{DFG}, German Research Foundation) through grant
396021762 - TRR~257 ``Particle Physics Phenomenology after the Higgs
Discovery'', through grant 513989149, Germany’s Excellence Strategy – EXC 3107 – Project 533766364,
\abbrev{UK STFC} grant ST/X000494/1, the Swiss National Science Foundation (\abbrev{SNSF}) under contract TMSGI2\_211209, the
National Science Foundation under grant PHY-2209185, and the \abbrev{DOE} Topical
Collaboration ``Nuclear Theory for New Physics'' award No.\ DE-SC0023663.

\paragraph{Data availability ---}
The data that support the findings of this article are openly available~\cite{DatasetGF}.

  \paragraph{Note added ---}
  While this Letter was under review, Ref.\,\cite{Moretti:2026waq} presented
  the \nnlo\ Wilson coefficients for heavy-meson lifetime ratios and combined
  them with the bag parameters determined here.  It finds that our lattice
  results significantly change the charm lifetime-ratio predictions compared
  to the \hqet{} sum-rule results~\cite{Black:2024bus}, reduce the associated
  uncertainties, and improve the agreement with experiment.

\bibliography{lit.bib}

\clearpage\newpage
\setcounter{page}{1}
\onecolumngrid
\section*{End Matter: Bag Parameters for Heavy Meson Lifetimes}
In this End Matter, we provide two tables with supporting information to this Letter.
\Cref{tab:confs} lists the set of gauge-field ensembles used in this study, specifying the relevant ensemble parameters as well as statistics used and minimum source separation $\Delta T$ considered in the analysis.
The final bag parameter results are presented in~\cref{tab:errorbudget} with a full error budget.

\begin{table}[h]
    \centering
    \begin{tabular}{l@{~~~}c@{~~~}c@{~~~}c@{~~~}c@{~~~}c@{~~~}cccccccc}
        \hline\hline & L/a & T/a & $a^{-1}[\gev]$ & $M_\pi[\mev]$ & $am_s^{\rm sea}$ & $am_s^\text{val}$ & $am_c$ 
                      & $N_\text{src}\times\text{N}_{\text{conf}}$ & $\sigma$ & $N_\sigma$ & $\Delta T_{\rm min}$ \\\hline\hline
        C1 & 24 & 64 & 1.7848(50) & 339.8(1.2) & 0.04\phantom{144} & 0.03224 & 0.64& $32\times101$ & 4.5 & 400 & 20 \\
        C2 & 24 & 64 & 1.7848(50) & 430.6(1.4) & 0.04\phantom{144} & 0.03224 & 0.64 & $32\times101$ & 4.5 & 400 & 20 \\[1.2ex]
        M1 & 32 & 64 & 2.3833(86) & 303.6(1.4) & 0.03\phantom{144} & 0.02477 & 0.45 & $32\times\phantom{0}79$ & 6.5 & 400 & 22 \\
        M2 & 32 & 64 & 2.3833(86) & 360.7(1.6) & 0.03\phantom{144} & 0.02477 & 0.45 & $32\times\phantom{0}89$ & 6.5 & 100 & 22 \\
        M3 & 32 & 64 & 2.3833(86) & 410.8(1.7) & 0.03\phantom{144} & 0.02477 & 0.45 & $32\times\phantom{0}68$ & 6.5 & 100 & 22 \\[1.2ex]
        F1S & 48 & 96 & 2.785(11) & 267.6(1.3) & 0.02144 & 0.02167 & 0.37 & $24\times\phantom{0}98$ & & & 28 
        \\\hline\hline 
    \end{tabular}
    \caption{RBC/UKQCD's $N_f=2+1$ Shamir \dwf\ ensembles with Iwasaki gauge action \cite{Allton:2008pn,Aoki:2010dy,Blum:2014tka,Boyle:2018knm} characterized by inverse lattice spacing ($a^{-1}$), unitary pion mass ($M_\pi$), valence strange ($am_s^\text{val}$) and charm ($am_c$) quark masses. We use coarse (C1,C2), medium (M1,M2,M3) and one fine ensemble (F1S). The Jacobi smearing procedure was applied on the C and M ensembles with smearing width $\sigma$ and $N_\sigma$ iterations. $\Delta T_{\rm min}$ indicates the minimum source separation $\Delta T$ considered to have sufficient resolution of the ground state.}
    \label{tab:confs}
\end{table}

\begin{table}[h]
    \centering
    \begin{tabular}{l|@{~~~}c@{~~~}|@{~~~}c@{~~~}|@{~~~}c@{~~~}|@{~~~}c@{~~~}} 
        \hline\hline 
        Bag parameter & $B_1^{\MSbar}$ & $B_2^{\MSbar}$ & $\epsilon_1^{\MSbar}$ & $\epsilon_2^{\MSbar}$ \\[3pt] \hline
        Central value at $\mu_0=3\gev$ & $1.0524$ & $0.9621$ & $-0.2275$ & $-0.0005$ \\[3pt]
        \hline
        \multicolumn{5}{c}{Error budget} \\[3pt] \hline
        Stat.~error and $\tau\to0$ extrapolation (`GF') & $0.0069$ & $0.0025$ & $\phantom{-}0.0041$ & $\phantom{-}0.0003$ \\
        Continuum limit (`CL') & $0.0001$ & $0.0018$ & $\phantom{-}0.0042$ & $\phantom{-}0.0000$\\
        Perturbative truncation (`PT') & $0.0014$ & $0.0013$ & $\phantom{-}0.0046$ & $\phantom{-}0.0007$ \\
        Other systematics (`OS') & $0.0067$ & $0.0062$ & $\phantom{-}0.0014$ & $\phantom{-}0.0000$ \\ \hline
        {\bf Total} & $0.0097$ & $0.0070$ & $\phantom{-}0.0076$ & $\phantom{-}0.0008$ \\ 
        \hline\hline
    \end{tabular}
    \caption{Values and error budgets for the $\Delta Q=0$ bag parameters in the \MSbar{} scheme for our choice of evanescent operators at $\mu_0=3\gev$. Individual errors are added in quadrature to obtain the total uncertainty. The `GF' error refers to the combined statistical and systematic uncertainty obtained from our set of fits performing the zero-flow-time extrapolation; `CL' is an estimate of systematic effects in our continuum limit by performing our analysis without the coarse C1 and C2 ensembles and taking half the difference of its zero-flow-time result with our central result (shown by the gray data points in~\cref{fig:lattice+sftx}); `PT' estimates the effect of truncating the perturbative series to \nnlo\ by taking half the difference between the \nnlo\ results (brown squares) and the \nlo\ results (pink circles); `OS' incorporates other systematic effects, specifically quark-mass tuning and residual chiral-symmetry breaking of the light quarks. For further details, see Ref.~\cite{Black:2026dzp}.}
    \label{tab:errorbudget}
\end{table}
\twocolumngrid

\end{document}